\documentclass[aps,prd,10pt,nofootinbib,notitlepage]{revtex4-1}
\usepackage{amsmath,amsthm,amssymb,bm}
\usepackage[dvipdfmx]{graphicx}
\usepackage[dvipdfmx]{color}

\def\be{\begin{equation}}
\def\ee{\end{equation}}
\def\ba{\begin{eqnarray}}
\def\ea{\end{eqnarray}}

\begin{document}

\title{Hidden symmetry and the separability of the Maxwell equation\\
on the Wahlquist spacetime}

\author{Tsuyoshi Houri$^1$, Norihiro Tanahashi$^2$, Yukinori Yasui$^3$}

\affiliation{
$^1$National Institute of Technology, Maizuru College, Kyoto 625-8511, Japan,\\
$^2$Institute of Mathematics for Industry, Kyushu University, Fukuoka 819-0395, Japan,\\
$^3$Institute for Fundamental Sciences, Setsunan University, Osaka 572-8508, Japan
}

\begin{abstract}
We examine hidden symmetry and its relation to the separability
of the Maxwell equation on the Wahlquist spacetime.
After seeing that the Wahlquist spacetime is a type-D spacetime
whose repeated principal null directions are
shear-free and geodesic, we show that the spacetime admits three gauged 
conformal Killing-Yano (GCKY) tensors which are in a relation
with torsional conformal Killing-Yano tensors.
As a by-product, we obtain an ordinary CKY tensor.
We also show that thanks to the GCKY tensors,
the Maxwell equation reduces to three Debye equations,
which are scalar-type equations,
and two of them can be solved by separation of variables.
\end{abstract}

\maketitle

\section{Introduction}
\label{sec:1}

Hidden symmetry of spacetime has played an important role
in the study of black hole physics.
In particular, conformal Killing-Yano symmetry,
known as hidden symmetry of the Kerr spacetime,
has received special attention
since it has been crucial to understanding the separability
of various equations on a curved spacetime.

On the Kerr spacetime, the Hamilton-Jacobi equation for geodesics,
the Klein-Gordon equation, and the Dirac equation
can be solved by separation of variables, and
these separabilities have been understood
in terms of a Killing-Yano tensor.
For electromagnetic and gravitational perturbations,
Teukolsky \cite{Teukolsky:1972,Teukolsky:1973} 
provided the scalar-type master equations,
which can be solved by separation of variables.
Cohen and Kegeles \cite{Cohen:1974} showed that, 
if a spacetime is algebraically special
and its repeated principal null direction (PND)
is shear-free and geodesic,\footnote{
These conditions are referred to as the generalized Goldberg-Sachs theorem
in \cite{Cohen:1974}.}
the Maxwell equation reduces to a scalar-type master equation
called the Debye equation.
They also showed that when their method is applied to the Kerr spacetime,
the Debye equation coincides with the Teukolsky equation.
Moreover, they extended their results to massless Dirac fields
on algebraically special spacetimes and gravitational perturbations
on vacuum spacetimes \cite{Cohen:1979}.
After a while, Benn, Charlton, and Kress \cite{Benn:1997}
unveiled the underlying structure of the works by Cohen and Kegeles
from the viewpoint of GCKY symmetry.
So far, while the GCKY symmetry has been related to 
obtaining the scalar-type master equations
from the Dirac, Maxwell, and linearized Einstein equations,
the separability of those equations has not been well understood.
To fill the gap is one of the aims in this paper.

In this paper, we examine hidden symmetry of the Wahlquist spacetime
\cite{Wahlquist:1968,Kramer:1985,Senovilla:1987,Wahlquist:1992,
Bradley:1999,Mars:2001,Hinoue:2014},
which is a stationary, axially symmetric solution to the Einstein equation
with rigidly rotating perfect fluids.
According to the Goldberg-Sachs theorem, the repeated PNDs on a type-D
vacuum spacetime are shear-free and geodesic.
However, since the Wahlquist metric is a non-vacuum type-D spacetime,
it is interesting to ask if the repeated PNDs are 
shear-free and geodesic.
According to \cite{Cohen:1974,Benn:1997}, if they are so,
we obtain GCKY tensors with a certain condition
and hence the Maxwell equation reduces to the Debye equation.
Since the Wahlquist spacetime is known to admit a
torsional conformal Killing-Yano (TCKY) tensor \cite{Hinoue:2014},
there may be some relation between the gauged and torsional CKY tensors, 
and it is another aim of this work to clarify it.
We also examine whether the Debye equation
can be solved by separation of variables.

\begin{figure}[t]
\centering
 \begin{minipage}{54mm}
  \begin{center}
   \includegraphics[width=52mm]{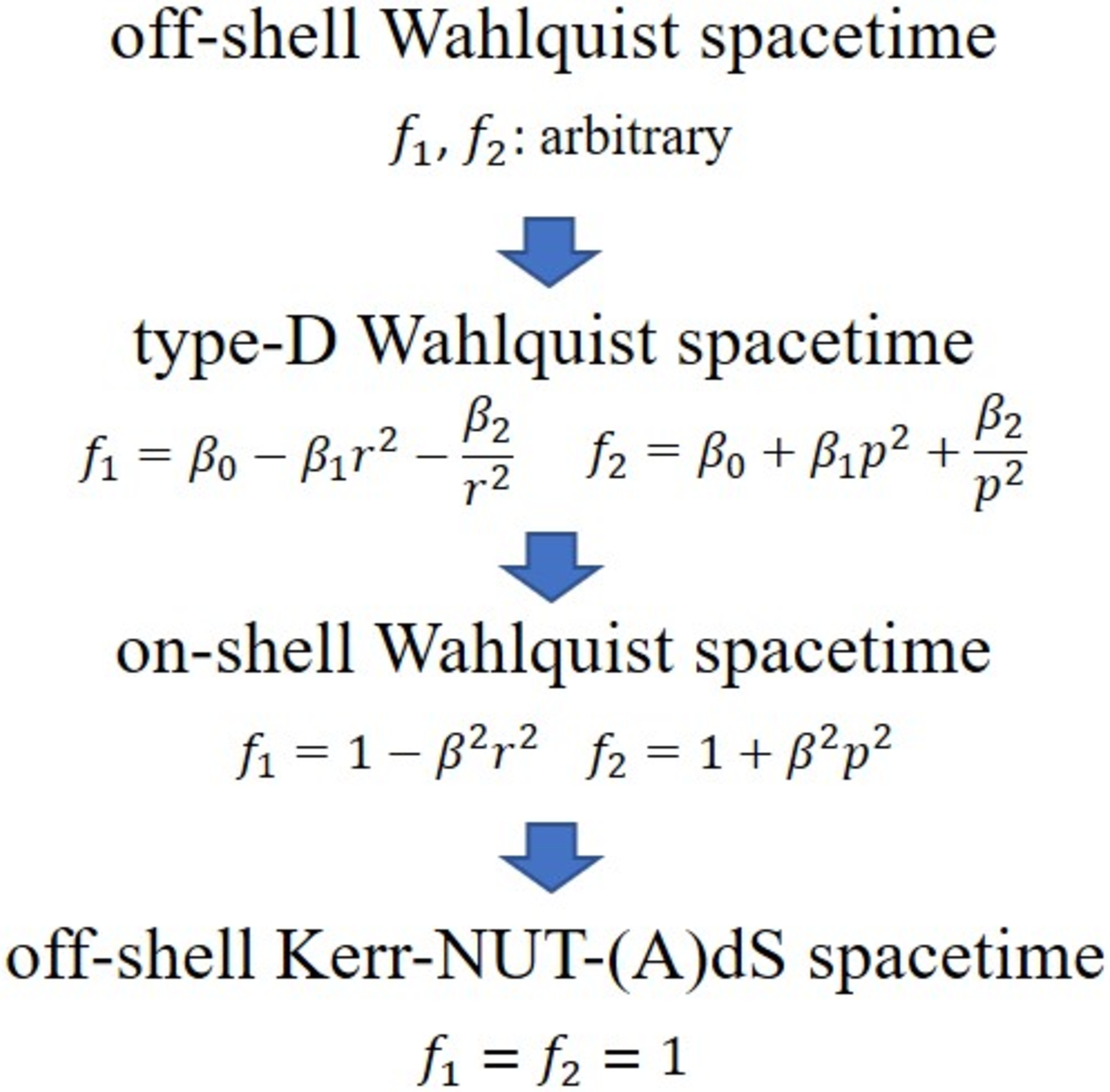}
  \end{center}
  \caption{Our terminology for spacetimes in this work,
  which depend on the functions $f_1$ and $f_2$.}
  \label{fig:terminology}
 \end{minipage}
 \qquad
 \begin{minipage}{105mm}
  \begin{center}
   \includegraphics[width=105mm]{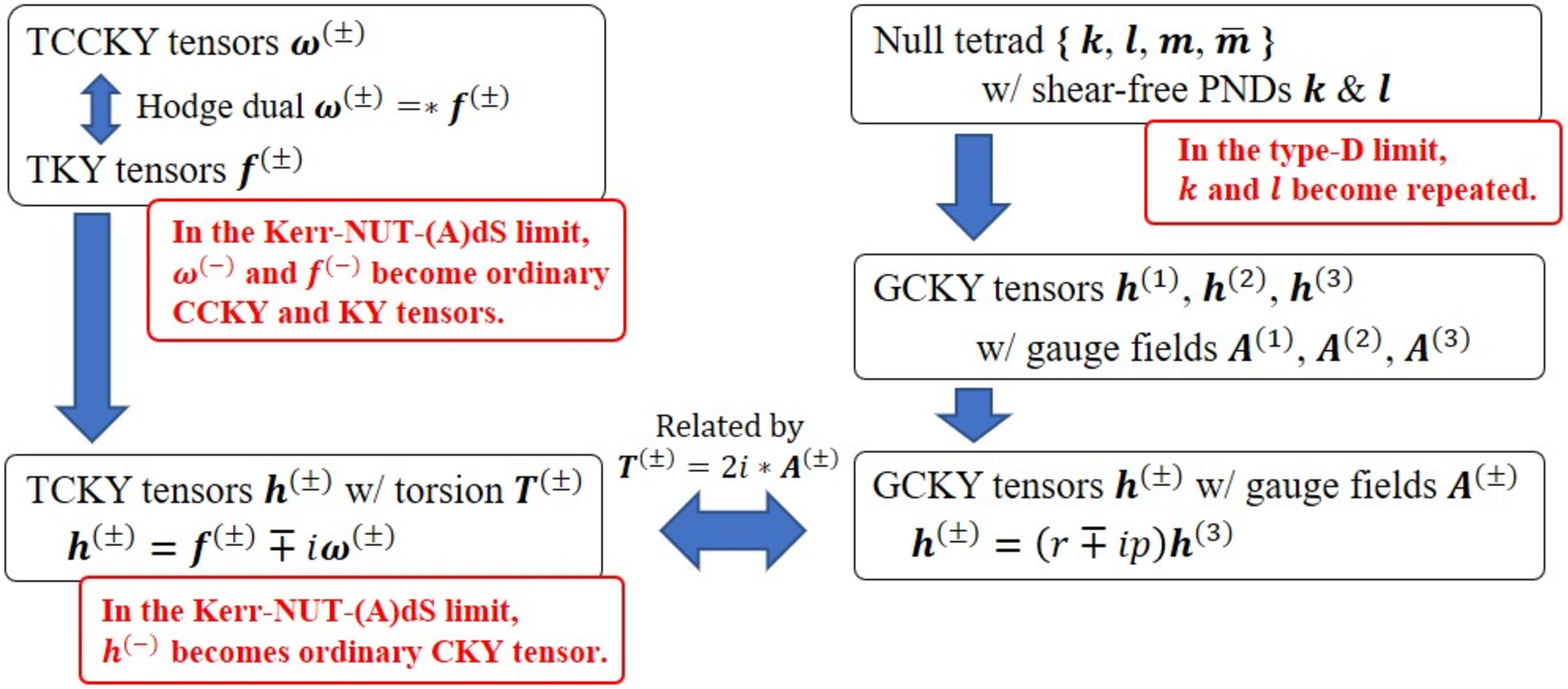}
  \end{center}
  \caption{Relations between hidden symmetry of the Wahlquist spacetime.
  The arrows show that, if $A\Rightarrow B$, 
  $B$ is induced or constructed from $A$.}
  \label{fig:relation}
 \end{minipage}
\end{figure}

This paper is organized as follows.
In the next section, we examine hidden symmetry of the Wahlquist spacetime.
In Sec.~\ref{sec:3}, we show that the Maxwell equation reduces to
three Debye equations, and two of them can be solved by separation of variables.
The last section is devoted to summary and discussion.
In appendix, we briefly show the separability of the Dirac equation
on the Wahlquist spacetime.
We summarize our terminology for spacetimes in FIG.~\ref{fig:terminology}
and relations between hidden symmetry of the Wahlqusit spacetime 
in FIG.~\ref{fig:relation}.

\noindent

\section{Hidden symmetry of the Wahlquist spacetime}
\label{sec:2}

In this section, we examine hidden symmetry
of the Wahlquist spacetime.
Although some results about the torsional conformal Killing-Yano tensor
on this spacetime were already found in \cite{Hinoue:2014},
the aim of this section is to elucidate the gauged conformal 
Killing-Yano tensors
and their relations to the torsional one.

\subsection{Off-shell  Wahlquist metric}

The Wahlquist metric (see, e.g. \cite{Hinoue:2014}) is written as 
\begin{align}
 ds^2 
 = -\frac{{\cal Q}}{r^2+p^2}(d\tau-p^2d\sigma)^2
   + \frac{r^2+p^2}{{\cal Q} f_1}dr^2
  + \frac{r^2+p^2}{{\cal P} f_2}dp^2
  +\frac{{\cal P}}{r^2+p^2}(d\tau+r^2d\sigma)^2 \,,
  \label{Wahlquistmetric}
\end{align}
where ${\cal Q}$ and ${\cal P}$ are given by
\begin{align}
 {\cal Q} =& Q_0 + a_2 r\sqrt{1-\beta^2r^2}+\nu_0 r^2
  +\frac{\mu_0}{\beta^2}\left[
 r^2-\frac{r\,\textrm{Arcsin}(\beta r)\sqrt{1-\beta^2r^2}}{\beta}
 \right] \,, \\
 {\cal P} =& Q_0 + a_1 p\sqrt{1+\beta^2p^2}-\nu_0 p^2
 -\frac{\mu_0}{\beta^2}\left[
 p^2-\frac{p\,\textrm{Arcsin}(\beta p)\sqrt{1+\beta^2p^2}}{\beta}
 \right] \,,
\end{align}
and $f_1$ and $f_2$ are given by
\be
 f_1 = 1-\beta^2r^2 \,, \qquad
 f_2 = 1+\beta^2p^2 \,,
 \label{Def_f1f2_0}
\ee
with constants $Q_0$, $a_1$, $a_2$, $\nu_0$, $\mu_0$ and $\beta$.

To handle a wider class of metrics including the Wahlquist metric,
we consider the metric \eqref{Wahlquistmetric}
with ${\cal Q}$, ${\cal P}$, $f_1$, and $f_2$
replaced with arbitrary functions of single variable,
${\cal Q}(r)$, ${\cal P}(p)$, $f_1(r)$, and $f_2(p)$
and call it the off-shell Wahlquist metric.
As a special class, the Wahlquist metric
with $\beta=0$ is called the Kerr-NUT-(A)dS metric.
Moreover, the off-shell Wahlquist metric with $f_1=f_2=1$
is called the off-shell Kerr-NUT-(A)dS metric.

\subsection{Torsional Killing-Yano tensor
and Killing-St\"ackel tensor}

In \cite{Hinoue:2014}, it was shown that the off-shell Wahlquist metric 
admits torsional Killing-Yano (TKY) tensors ${\bm f}^{(\pm)}$ 
and torsional closed conformal Killing-Yano (TCCKY) tensors ${\bm \omega}^{(\pm)}$, 
defined by the equations
\begin{align}
 \nabla^{(T)}_a f_{bc} &= \nabla^{(T)}_{[a}f_{bc]} \,, \\
 \nabla^{(T)}_a \omega_{bc} &= g_{ab}\xi^{(T)}_c - g_{ac}\xi^{(T)}_b \,,
\end{align}
where
\begin{align}
 \xi^{(T)}_a = \frac{1}{3}\nabla^{(T)b}\omega_{ba} \,.
\end{align}
Here, $\nabla^{(T)}_a$ is the covariant derivative
with totally antisymmetric torsion ${\bm T}$
and it acts on a 2-form ${\bm \Psi}$ as
\begin{align}
 \nabla^{(T)}_a \Psi_{bc} 
= \nabla_a \Psi_{bc} + \frac{1}{2}T_{ab}{}^d\Psi_{cd}
  -\frac{1}{2}T_{ac}{}^d\Psi_{bd} \,.
\end{align}
To describe such symmetries,
it is convenient to introduce the orthonormal basis
$\{{\bm e}^\mu\}$ as
\begin{align}
 {\bm e}^0 
 = \sqrt{\frac{{\cal Q}}{r^2+p^2}}(d\tau-p^2d\sigma) \,, \quad
 {\bm e}^1
 = \sqrt{\frac{r^2+p^2}{{\cal Q}f_1}}dr \,, \quad
 {\bm e}^2
 = \sqrt{\frac{r^2+p^2}{{\cal P}f_2}}dp \,, \quad
 {\bm e}^3
 = \sqrt{\frac{{\cal P}}{r^2+p^2}}(d\tau+r^2d\sigma) \,,
\end{align}
with which the metric is given by
\be
 {\bm g} = -{\bm e}^0{\bm e}^0
 + {\bm e}^1{\bm e}^1
 + {\bm e}^2{\bm e}^2
 + {\bm e}^3{\bm e}^3 \,.
\ee
In terms of these orthonormal basis,
the TKY tensors ${\bm f}^{(\pm)}$ 
and as their Hodge dual the TCCKY tensors 
${\bm \omega}^{(\pm)}=*{\bm f}^{(\pm)}$ are given by
\begin{align}
&{\bm f}^{(\pm)} = p\,{\bm e}^0\wedge{\bm e}^1
 \pm r\,{\bm e}^2\wedge{\bm e}^3 \,, \label{Def_f} \\
&{\bm \omega}^{(\pm)} = p\,{\bm e}^2\wedge{\bm e}^3
 \mp r\,{\bm e}^0\wedge{\bm e}^1 \,, \label{Def_omega}
\end{align}
with the common totally antisymmetric torsion
\begin{align}
 {\bm T}^{(\pm)} =&
 \frac{2r(\sqrt{f_1}\pm \sqrt{f_2})}{r^2+p^2}
 \sqrt{\frac{{\cal P}}{r^2+p^2}}
 \,{\bm e}^0\wedge {\bm e}^1\wedge {\bm e}^3
 \pm \frac{2p(\sqrt{f_1}\pm \sqrt{f_2})}{r^2+p^2}
 \sqrt{\frac{{\cal Q}}{r^2+p^2}}
 \,{\bm e}^0\wedge {\bm e}^2\wedge {\bm e}^3 \,.
 \label{Def_T}
\end{align}
It turns out that ${\bm T}^{(-)}$ vanishes
in the limit to the off-shell Kerr-NUT-(A)dS metric,
and then ${\bm f}^{(-)}$ and ${\bm \omega}^{(-)}$
become ordinary KY and CCKY tensors.
On the other hand, ${\bm T}^{(+)}$ does not vanish
in the same limit.

The squares of the TKY and TCCKY tensors, 
defined by $K_{ab}=f^{(\pm)}_{ac}f^{(\pm)}{}_b{}^c$ 
and $Q_{ab}=\omega^{(\pm)}_{ac}\omega^{(\pm)}{}_b{}^c$
with double sign correspond, 
give rise to a Killing-St\"ackel tensor ${\bm K}$
and a conformal Killing-St\"ackel tensor ${\bm Q}$,
defined by
\begin{align}
 \nabla_{(a}K_{bc)} = 0 \,, \quad
 \nabla_{(a}Q_{bc)} = q_{(a}g_{bc)} \,.
\end{align}
Thus, we obtain
\begin{align}
 {\bm K} &= p^2({\bm e}^0{\bm e}^0-{\bm e}^1{\bm e}^1)
            +r^2({\bm e}^2{\bm e}^2+{\bm e}^3{\bm e}^3) \,, \\
 {\bm Q} &= r^2({\bm e}^0{\bm e}^0-{\bm e}^1{\bm e}^1)
            +p^2({\bm e}^2{\bm e}^2+{\bm e}^3{\bm e}^3)\,.
\end{align}
These tensors are known to guarantee 
the existence of conserved quantities
along geodesics. Thanks to them,
the Hamilton-Jacobi equation for geodesics
on the Wahlquist spacetime is completely integrable
in the Liouville sense.

Furthermore, a linear combination of TKY and TCCKY tensors
with a common totally antisymmetric torsion
gives rise to a TCKY tensor ${\bm h}$, defined by
\begin{align}
 \nabla^{(T)}_a h_{bc} &= \nabla^{(T)}_{[a}h_{bc]} 
 + g_{ab}\xi^{(T)}_c - g_{ac}\xi^{(T)}_b \,,
 \label{CKYTeq}
\end{align}
where
\begin{align}
 \xi^{(T)}_a = \frac{1}{3}\nabla^{(T)b}h_{ba} \,.
\end{align}
Hence, we are able to construct two complex TCKY tensors,
\begin{align}
& {\bm h}^{(+)}
 = {\bm f}^{(+)} - i {\bm \omega}^{(+)}
 = -(r - i p)\left({\bm e}^0\wedge{\bm e}^1
   - i \,{\bm e}^2\wedge{\bm e}^3\right) \,, 
   \label{Def_hplus} \\
& {\bm h}^{(-)}
  = {\bm f}^{(-)} + i {\bm \omega}^{(-)}
  = -(r +i p)\left({\bm e}^0\wedge{\bm e}^1
   - i \,{\bm e}^2\wedge{\bm e}^3\right) \,,
   \label{Def_hminus}
\end{align}
where ${\bm h}^{(\pm)}$ are TCKY tensors
with the antisymmetric torsions ${\bm T}^{(\pm)}$,
given by (\ref{Def_T}), corresponding to the signs.
As shown later, these TCKY tensors are also GCKY tensors.
Moreover, we will see that, when we consider the on-shell
Wahlquist spacetime, ${\bm h}^{(\pm)}$ are
obtained from an ordinary CKY tensor by certain gauge transformations.

\subsection{Principal null directions}

In order to see if the PNDs in the Wahlquist spacetime
are shear-free and geodesic,
it is convenient to introduce a null tetrad
$\{{\bm k},{\bm l},{\bm m},\bar{{\bm m}}\}$
and calculate the corresponding spin coefficients.
If $\kappa$, $\sigma$, $\lambda$, and $\mu$ vanish,
${\bm k}$ and ${\bm l}$ are shear-free and geodesic.

To do so, we first introduce the orthonormal
vector basis $\{{\bm e}_\mu\}$
by ${\bm e}^\mu({\bm e}_\nu)=\delta^\mu_\nu$, which are given by
\begin{align}
&{\bm e}_0 
 = \frac{1}{\sqrt{{\cal Q}(r^2+p^2)}}(r^2\partial_\tau-\partial_\sigma) \,, \quad
 {\bm e}_1
 = \sqrt{\frac{{\cal Q}f_1}{r^2+p^2}}\partial_r \,, \quad
 {\bm e}_2
 = \sqrt{\frac{{\cal P}f_2}{r^2+p^2}}\partial_p \,, \quad
 {\bm e}_3
 = \frac{1}{\sqrt{{\cal P}(r^2+p^2)}}(p^2\partial_\tau+\partial_\sigma) \,.
\end{align}
Using these orthonormal basis, we introduce the null basis
$\{{\bm k},{\bm l},{\bm m},\bar{{\bm m}}\}$ by
\begin{align}
 {\bm k} 
 = \sqrt{\frac{r^2+p^2}{{\cal Q}f_1}}
 ({\bm e}_0+{\bm e}_1) \,, \quad
 {\bm l}
 = \frac{1}{2}\sqrt{\frac{{\cal Q}f_1}{r^2+p^2}}
 ({\bm e}_0-{\bm e}_1) \,, \quad
 {\bm m}
 = \frac{\sqrt{r^2+p^2}}{\sqrt{2}\bar{\chi}}
 ({\bm e}_2-i{\bm e}_3)\,. \label{NB}
\end{align}
where
\begin{align}
 \chi = r+ip \,, \quad
 \bar{\chi} = r-ip \,,
\end{align}
and $\bar{{\bm m}}$ is the complex conjugate to ${\bm m}$.

Using these null basis, we calculate 
the Weyl scalars $\{\Psi_0,\Psi_1,\Psi_2,\Psi_3,\Psi_4\}$.\footnote{
The Weyl scalars are defined by 
$\Psi_0=-{\bm W}({\bm k},{\bm m},{\bm k},{\bm m})$,
$\Psi_1=-{\bm W}({\bm k},{\bm l},{\bm k},{\bm m})$,
$\Psi_2=-{\bm W}({\bm k},{\bm m},\bar{{\bm m}},{\bm l})$,
$\Psi_3=-{\bm W}({\bm k},{\bm l},\bar{{\bm m}},{\bm l})$, 
and $\Psi_4=-{\bm W}({\bm l},\bar{{\bm m}},{\bm l},\bar{{\bm m}})$.}
It turns out that, since $\Psi_0$ and $\Psi_4$ are vanishing
for arbitrary $f_1$ and $f_2$, ${\bm k}$ and ${\bm l}$ are PNDs.
Moreover, if and only if $f_1$ and $f_2$ are given by
\be
 f_1 = \beta_0-\beta_1 r^2-\frac{\beta_2}{r^2} \,, \quad
 f_2 = \beta_0+\beta_1 p^2+\frac{\beta_2}{p^2} \,,
 \label{Def_f1f2}
\ee
with constants $\beta_0$, $\beta_1$, $\beta_2$,
all the Weyl scalars other than $\Psi_2$ vanish,
so that ${\bm k}$ and ${\bm l}$ become repeated PNDs.
This means that the Wahlquist metric
with $f_1$ and $f_2$ given by \eqref{Def_f1f2}
is of type D in the Petrov classification.
Hereafter, we call it the type-D Wahlquist metric.
The original Wahlquist metric is a particular case
of the type-D Wahlquist metric
with $\beta_0=1$, $\beta_1=\beta^2$, and $\beta_2=0$.

For arbitrary $f_1$ and $f_2$,
the spin coefficients\footnote{
The spin coefficients are defined by
$\kappa=(\nabla_{\bm k}{\bm k},{\bm m})$, 
$\sigma=(\nabla_{\bm m}{\bm k},{\bm m})$, 
$\lambda=(\nabla_{\bar{{\bm m}}}\bar{{\bm m}},{\bm l})$, 
$\nu=(\nabla_{\bm l}\bar{{\bm m}},{\bm l})$,
$\rho=(\nabla_{\bar{{\bm m}}}{\bm k},{\bm m})$,
$\mu=(\nabla_{\bm m}\bar{{\bm m}},{\bm l})$,
$\tau=(\nabla_{\bm l}{\bm k},{\bm m})$, 
$\pi=(\nabla_{\bm k}\bar{{\bm m}},{\bm l})$, 
$\epsilon=\frac{1}{2}[(\nabla_{\bm k}{\bm k},{\bm l})
+(\nabla_{\bm k}\bar{{\bm m}},{\bm m})]$, 
$\gamma=\frac{1}{2}[(\nabla_{\bm l}{\bm k},{\bm l})
+(\nabla_{\bm l}\bar{{\bm m}},{\bm m})]$, 
$\alpha=\frac{1}{2}[(\nabla_{\bar{{\bm m}}}{\bm k},{\bm l})
+(\nabla_{\bar{{\bm m}}}\bar{{\bm m}},{\bm m})]$, and 
$\beta=\frac{1}{2}[(\nabla_{\bm m}{\bm k},{\bm l})
+(\nabla_{\bm m}\bar{{\bm m}},{\bm m})]$.
} are given by
\begin{align}
&\kappa = \sigma = \lambda = \nu = 0 \,, \quad
 \epsilon = \frac{ip(\sqrt{f_1}
-\sqrt{f_2})}{2(r^2+p^2)\sqrt{f_1}}+\frac{f_1'}{4f_1} \,, \quad
 \gamma = \frac{{\cal Q}f_1}{r^2+p^2}
\left(\frac{(2r-ip)\sqrt{f_1}-ip\sqrt{f_2}}{4(r^2+p^2)\sqrt{f_1}}
-\frac{f_1'}{8f_1}
-\frac{{\cal Q}'}{4{\cal Q}}\right)\,, \nonumber\\
&\rho = \frac{r\sqrt{f_1}-ip\sqrt{f_2}}{(r^2+p^2)\sqrt{f_1}} \,, \quad
 \mu = \frac{(r\sqrt{f_1}-ip\sqrt{f_2}){\cal Q}\sqrt{f_1}}{2(r^2+p^2)^2}\,, \quad
 \tau = \frac{i(r\sqrt{f_1}-ip\sqrt{f_2})\sqrt{{\cal P}}}{\sqrt{2}\bar{\chi}(r^2+p^2)} \,, \quad
 \pi = -\frac{i(r\sqrt{f_1}-ip\sqrt{f_2})\sqrt{{\cal P}}}{\sqrt{2}\chi(r^2+p^2)} \,, \nonumber\\
&\alpha = -\frac{\sqrt{{\cal P}f_2}}{\sqrt{2}\chi}
\left(\frac{i(r\sqrt{f_1}+(r-2ip)\sqrt{f_2})}{2(r^2+p^2)\sqrt{f_2}}
-\frac{{\cal P}'}{4{\cal P}}\right) \,, \quad
 \beta = \frac{\sqrt{{\cal P}f_2}}{\sqrt{2}\bar{\chi}}
\left(\frac{ir(\sqrt{f_1}-\sqrt{f_2})}{2(r^2+p^2)\sqrt{f_2}}
-\frac{{\cal P}'}{4{\cal P}}\right) \,,
\label{spincoefficients}
\end{align}
which show that ${\bm k}$ and ${\bm l}$ are shear-free and geodesic
for arbitrary $f_1$ and $f_2$.

For later calculation, we introduce the null 1-form basis
$\{{\bm k}_*,{\bm l}_*,{\bm m}_*,\bar{{\bm m}}_*\}$,
which are dual to (\ref{NB}) in the sense that 
for the null vector basis 
$\{{\bm E}_a\}=\{{\bm k},{\bm l},{\bm m},\bar{{\bm m}}\}$,
the dual basis 
$\{{\bm E}^a_*\}=\{{\bm k}_*,{\bm l}_*,{\bm m}_*,\bar{{\bm m}}_*\}$
are given by ${\bm E}^a_*({\bm E}_b)=\delta^a_b$.
The explicit forms are given by
\begin{align}
 {\bm k}_*
 = \frac{1}{2}\sqrt{\frac{{\cal Q}f_1}{r^2+p^2}}
 ({\bm e}^0+{\bm e}^1) \,, \quad
 {\bm l}_*
 = \sqrt{\frac{r^2+p^2}{{\cal Q}f_1}}
 ({\bm e}^0-{\bm e}^1) \,, \quad
 {\bm m}_*
 = \frac{\sqrt{r^2+p^2}}{\sqrt{2}\chi}
 ({\bm e}^2+i{\bm e}^3) \,, \label{NB_flat}
\end{align}
and $\bar{{\bm m}}_*$ is the complex conjugate to ${\bm m}_*$.

\subsection{Gauged conformal Killing-Yano tensors}

According to \cite{Benn:1997}, a shear-free PND provides 
a GCKY tensor ${\bm h}$, defined by
\begin{align}
 \nabla^{(A)}_a h_{bc} &= \nabla^{(A)}_{[a}h_{bc]} 
 + g_{ab}\xi^{(A)}_c - g_{ac}\xi^{(A)}_b \,,
 \label{GCKYeq}
\end{align}
where
\begin{align}
 \xi^{(A)}_a = \frac{1}{3}\nabla^{(A)b}h_{ba} \,.
\end{align}
Here, $\nabla^{(A)}_a$ is the gauge-covariant derivative
with gauge field ${\bm A}$ given by $\nabla^{(A)}_a
= \nabla_a + A_a$. We note that this gauge-covariant derivative
acts on a $p$-form $\Psi_{a_1\dots a_p}$ as
\begin{align}
 \nabla^{(A)}_{a}\Psi_{b_1\dots b_p}
 = (\nabla_a+A_a)\Psi_{b_1\dots b_p} \,.
\end{align}

For the off-shell Wahlquist metric,
we can construct three GCKYs
\begin{align}
 \bm{h}^{(1)} &= \bm{l}_* \wedge \bar{\bm{m}}_* \,, 
 \label{Def_h1}\\
 \bm{h}^{(2)} &= \bm{k}_* \wedge \bm{m}_* \,, 
 \label{Def_h2}\\
 \bm{h}^{(3)} &= \bm{k}_* \wedge \bm{l}_* 
 - \bm{m}_* \wedge \bar{\bm{m}}_* \,,
 \label{Def_h3}
\end{align}
with the gauge fields
\begin{align}
 \bm{A}^{(1)} =& 2(\epsilon + \rho)\bm{k}_* + 2\gamma \bm{l}_* 
 + 2(\beta + \tau) \bm{m}_* + 2\alpha \bar{\bm{m}}_* \,, \label{gp_gcky_1}\\
 \bm{A}^{(2)} =& -2\epsilon\bm{k}_* -2(\gamma+\mu) \bm{l}_*
   -2\beta  \bm{m}_* -2(\alpha+\pi)  \bar{\bm{m}}_* \,, \label{gp_gcky_2}\\
 \bm{A}^{(3)} =& \rho \bm{k}_* -\mu \bm{l}_*
  + \tau \bm{m}_* -\pi  \bar{\bm{m}}_* \,.
  \label{gp_gcky_3}
\end{align}
These gauge fields satisfy the relation
\begin{equation}
 \bm{A}^{(1)} + \bm{A}^{(2)} = 2\bm{A}^{(3)} \,.
 \label{A-relation}
\end{equation}
Particularly, it is remarkable that if, and only if, $f_1$ and $f_2$
are given by (\ref{Def_f1f2}), the field strength
${\bm F}^{(3)}=d{\bm A}^{(3)}$ vanishes and hence ${\bm A}^{(3)}$
is given by ${\bm A}^{(3)}=d \log W$ with
\begin{align}
 W =& \Bigg(\left(\frac{r\sqrt{f_1}-ip\sqrt{f_2}}{r^2+p^2}\right)^2
 +\beta_1\Bigg)^{-1/2} \,.
 \label{Def_W}
\end{align}

In what follows, we show that
the TCKY tensors ${\bm h}^{(\pm)}$ obtained in \eqref{Def_hplus}
and \eqref{Def_hminus} are also GCKY tensors.
To show this, it is important that
the GCKY equation \eqref{GCKYeq} is covariant
under a gauge transformation
\be
 {\bm h}\to \tilde{{\bm h}}=\Omega {\bm h} \,,\quad
 {\bm A}\to \tilde{{\bm A}}={\bm A}- d\log \Omega \,,
 \label{GaugeTransf}
\ee
where $\Omega$ is some function.
Since ${\bm h}^{(3)}$ and ${\bm A}^{(3)}$ are written
in the orthonormal basis $\{{\bm e}^\mu\}$ as
\begin{align}
 {\bm h}^{(3)} 
 =& -{\bm e}^0\wedge {\bm e}^1 
 + i{\bm e}^2\wedge {\bm e}^3 \,, \\
 {\bm A}^{(3)}
 =& \frac{r\sqrt{f_1}-ip\sqrt{f_2}}{r^2+p^2}
 \sqrt{\frac{{\cal Q}}{r^2+p^2}}\,{\bm e}^1
 + \frac{ir\sqrt{f_1}+p\sqrt{f_2}}{r^2+p^2}
 \sqrt{\frac{{\cal P}}{r^2+p^2}}\,{\bm e}^2 \,,
\end{align}
we notice that
\begin{align}
 {\bm h}^{(\pm)} = (r\mp ip){\bm h}^{(3)} \,.
 \label{gaugetransf_h3tohpm}
\end{align}
This shows that the TCKY tensors ${\bm h}^{(\pm)}$ are obtained 
from the GCKY tensor ${\bm h}^{(3)}$ by the gauge transformation
\eqref{GaugeTransf} with $\Omega = (r\mp ip)$,
and hence ${\bm h}^{(\pm)}$ are GCKY tensors with
the gauge fields
\begin{align}
 {\bm A}^{(\pm)} =&
 \frac{ir(\sqrt{f_1}\pm \sqrt{f_2})}{r^2+p^2}
 \sqrt{\frac{{\cal P}}{r^2+p^2}}
 \,{\bm e}^2
 \mp \frac{ip(\sqrt{f_1}\pm \sqrt{f_2})}{r^2+p^2}
 \sqrt{\frac{{\cal Q}}{r^2+p^2}}
 \,{\bm e}^1 \,.
 \label{Def_Apm}
\end{align}
Compared with \eqref{Def_T}, these gauge fields ${\bm A}^{(\pm)}$
are related with the torsions ${\bm T}^{(\pm)}$ by the Hodge dual as
\begin{align}
 \bm{T}^{(\pm)} = 2i * {\bm A}^{(\pm)} \,.
\end{align}

\subsection{Conformal Killing-Yano tensor}

In the previous subsection, we found that 
on the type-D Wahlquist spacetime,
the gauge field ${\bm A}^{(3)}$ is closed.
If a gauge field ${\bm A}$ of a GCKY tensor ${\bm h}$ is closed,
we can erase it by performing a gauge transformation \eqref{GaugeTransf},
and then the GCKY tensor becomes an ordinary CKY tensor.
In fact, performing the gauge transformation \eqref{GaugeTransf}
with the function (\ref{Def_W}),
we can erase the gauge field ${\bm A}^{(3)}$,
and obtain a CKY tensor $\tilde{{\bm h}}^{(3)}$ as
\begin{align}
 \tilde{{\bm h}}^{(3)} = W {\bm h}^{(3)}  \,.
 \label{gtransf_h3}
\end{align}
To the best of our knowledge, this CKY tensor
on the Wahlquist spacetime is derived for the first time by this work.

Finally, we comment that
the CKY tensor $\tilde{{\bm h}}^{(3)}$ is related
with the TCKY tensors ${\bm h}^{(\pm)}$ by
\begin{align}
 \tilde{{\bm h}}^{(3)} = \frac{W}{r\mp ip}{\bm h}^{(\pm)} \,.
\end{align}
Particularly, in the limit to the Kerr-NUT-(A)dS spacetime,
we have $W=r+ip$, and hence the CKY tensor $\tilde{{\bm h}}^{(3)}$
coincides with ${\bm h}^{(-)}$.

\section{Separability of the Maxwell equation}
\label{sec:3}

In this section, we consider the Maxwell equation
on the Wahlquist spacetime.
We first briefly review the work by Benn, Charlton, and Kress \cite{Benn:1997}. Then, we apply it to the Maxwell equations on the off-shell and type-D 
Wahlquist spacetimes.

\subsection{Maxwell field, GCKY tensor, and Debye potential}
\label{sec:3-1}

We consider the Maxwell equation,
\begin{align}
 \nabla^a {\cal F}_{ab} = 0 \,,
 \label{Maxwelleq}
\end{align}
where ${\cal F}_{ab}=\nabla_a {\cal A}_b 
-\nabla_b {\cal A}_a$ is a Maxwell field.
Here, we have written the gauge potential of Maxwell field
in calligraphic letter to distinguish it from gauge fields of GCKY tensors.

Benn, Charlton, and Kress \cite{Benn:1997} showed that,
if a spacetime admits a GCKY tensor ${\bm h}$ with a gauge field ${\bm A}$,
given by \eqref{GCKYeq}, and it satisfies the eigenequation,
\begin{align}
 \frac{1}{2}C_{ab}{}^{cd}h_{cd}
 -\frac{1}{6}Rh_{ab}-F^c{}_{[a}h_{b]c}
  = \lambda h_{ab}
  \label{EigenEquation}
\end{align}
with some function $\lambda$,
where $C_{abcd}$ and $R$ are the Weyl and scalar curvatures,
and ${\bm F}$ is the field strength of ${\bm A}$,
i.e., ${\bm F}=d{\bm A}$,
then the Maxwell equation \eqref{Maxwelleq} reduces
to the scalar-type equation, 
\begin{align}
 g^{ab}(\nabla_a-A_a)(\nabla_b-A_b)\Phi 
 + \lambda\Phi = 0 \,.
 \label{Debyeeq}
\end{align}
This equation is called the Debye equation,
and its solution is called a Debye potential.
Given a Debye potential $\Phi$, we can reconstruct
the gauge potential $\bm{{\cal A}}$ of Maxwell field as
\begin{align}
 {\cal A}_a = - \nabla^b (\Phi h_{ba}) 
 + \frac{2}{3}\Phi (\nabla^b + A^b) h_{ba} \,.
 \label{gp_reconstruction}
\end{align}

It was shown in \cite{Houri:2019} that,
when we apply this formulation to the Maxwell equation
on the Kerr-NUT-(A)dS spacetime,
we obtain three Debye equations and two of them become separable.
The separable Debye equations reduce to the Teukolsky equations
\cite{Teukolsky:1972,Teukolsky:1972} with $s=\pm 1$
in the limit to the Kerr spacetime.
In the next subsection, we apply it to the Maxwell equation
on the Wahlquist spacetime.

It is worth stressing that the Debye equation \eqref{Debyeeq}
is covariant under the gauge transformation
\begin{align}
 \tilde{{\bm h}} = \Omega{\bm h} \,, \quad
 \tilde{{\bm A}} = {\bm A} - d\log \Omega \,, \quad
 \tilde{\Phi} = \Omega^{-1}\Phi \,.
 \label{gt}
\end{align}
In fact, after the gauge transformation, we have
\begin{align}
 \left[g^{ab}(\nabla_a-\tilde{A}_a)(\nabla_b-\tilde{A}_b)
 + \lambda\right] \tilde{\Phi} = 0 \,.
\end{align}
The gauge potential \eqref{gp_reconstruction} of Maxwell field
is gauge invariant, i.e.,
\begin{align}
 \tilde{\bm{{\cal A}}~} = \bm{{\cal A}} \,.
 \label{ginvariance_A}
\end{align}

\subsection{Debye equations}
\label{sec:3-2}

For the type-D Wahlquist spacetime,
we can confirm that the GCKY tensors $\bm{h}^{(i)}$,
given by \eqref{Def_h1}--\eqref{Def_h3}, satisfy
the eigenequation (\ref{EigenEquation}) with
\begin{align}
\lambda &= -\frac{R}{6}-4\Psi_2 \quad 
\textrm{for both ${\bm h}^{(1)}$ and ${\bm h}^{(2)}$} \,,\\
\lambda &= -\frac{R}{6}+2\Psi_2 \quad 
\textrm{for ${\bm h}^{(3)}$} \,,
\label{ev_gcky}
\end{align}
where $R$ is the scalar curvature, and $\Psi_2$ is the Weyl scalar.
Here, we note that the eigenequation for ${\bm h}^{(3)}$ is satisfied
with $f_1$ and $f_2$ arbitrary.

\subsubsection{Off-shell case}

When the metric is off-shell, that is,
$f_1$ and $f_2$ are arbitrary,
the only ${\bm h}^{(3)}$ satisfies 
the eigenequation \eqref{EigenEquation}.
Hence, we obtain the Debye equation for ${\bm h}^{(3)}$ as
\begin{align}
 \left[g^{ab}\left(\nabla_a-A^{(3)}_a\right)\left(\nabla_b-A^{(3)}_b\right)
 -\frac{R}{6}+2\Psi_2\right]\Phi = 0 \,,
\end{align}
which is explicitly given by\footnote{
We notice that the potential term (not including differentials)
is vanishing, which provides us that
\begin{align}
 \nabla^aA^{(3)}_a  = A^{(3)a}A^{(3)}_a 
 - \frac{R}{6} + 2\Psi_2 \,.
\end{align}}
\begin{align}
 \frac{1}{r^2+p^2}
 \Bigg[f_1 {\cal Q}\partial_r^2 
 +\left( f_1{\cal Q}' + \left(\frac{f_1'}{2}
 -\frac{2(r\sqrt{f_1}-ip\sqrt{f_2})}{(r^2+p^2)\sqrt{f_1}}\right)
 {\cal Q}\right)\partial_r 
 - \frac{1}{{\cal Q}}(r^2 \partial_\tau - \partial_\sigma)^2& \nonumber\\
 +f_2{\cal P}\partial_p^2
 + \left(f_2{\cal P}'+ \left(\frac{f_2'}{2}
 -\frac{2i(r\sqrt{f_1}-ip\sqrt{f_2})}{(r^2+p^2)\sqrt{f_2}}\right)
 {\cal P}\right)\partial_p
 + \frac{1}{{\cal P}}(p^2 \partial_\tau + \partial_\sigma)^2
  \Bigg]\Phi &= 0 \,.
  \label{Debye_3}
\end{align}
This equation cannot be separated even 
using the gauge transformation~(\ref{gt}).

Since ${\bm h}^{(1)}$ and ${\bm h}^{(2)}$ do not satisfy
the eigenequation \eqref{EigenEquation},
we cannot obtain the Debye equations for them.

\subsubsection{Type-D case}

Let us consider the type-D case,
that is, $f_1$ and $f_2$ are given by \eqref{Def_f1f2}.
In this case, since ${\bm h}^{(1)}$ and ${\bm h}^{(2)}$ satisfy
the eigenequation \eqref{EigenEquation},
we obtain three Debye equations for ${\bm h}^{(1)}$,
${\bm h}^{(2)}$, and ${\bm h}^{(3)}$.
However, none of them is separable.

To obtain the Debye equations that are separable,
we need to perform the gauge transformations
\begin{align}
&\tilde{{\bm h}}^{(1)} = \frac{W}{\chi} {\bm h}^{(1)} \,, \quad
 \tilde{{\bm A}}^{(1)} = {\bm A}^{(1)} - d \log \frac{W}{\chi} \,, 
 \label{gtransf_h1}\\
&\tilde{{\bm h}}^{(2)} = \chi W {\bm h}^{(2)} \,, \quad
 \tilde{{\bm A}}^{(2)} = {\bm A}^{(2)} - d \log (\chi W) \,,
 \label{gtransf_h2}
\end{align}
where $W$ is given by \eqref{Def_W}.
We note that $\tilde{{\bm A}}^{(1)} + \tilde{{\bm A}}^{(2)} = 0$
follows from \eqref{A-relation} and ${\bm A}^{(3)}=d \log W$,
and hence $\tilde{{\bm A}}^{(2)}=-\tilde{{\bm A}}^{(1)}$.
For these GCKY tensors,
the Debye equations are written as
\begin{align}
 \left[g^{ab}\left(\nabla_a+s \tilde{A}^{(1)}_a\right)
 \left(\nabla_b+s \tilde{A}^{(1)}_b\right)
 -\frac{R}{6} - 4s^2\Psi_2 \right]\tilde{\Phi} = 0 \,,
 \label{Debyeeq_s0}
\end{align}
where $s=-1$ for $\tilde{{\bm h}}^{(1)}$
and $s=1$ for $\tilde{{\bm h}}^{(2)}$.
The equation is explicitly given by
\begin{align}
&\frac{1}{r^2+p^2}
 \Bigg[f_1 {\cal Q}\partial_r^2 
 +\left( (s+1)f_1{\cal Q}' +(2s+1) \frac{f_1'}{2}{\cal Q}\right)\partial_r
 +f_2{\cal P}\partial_p^2
 + \left(f_2{\cal P}'+\frac{f_2'}{2}{\cal P}\right)\partial_p
 \nonumber\\
&\qquad \qquad 
  - \frac{1}{{\cal Q}}(r^2 \partial_\tau - \partial_\sigma)^2
  +s \sqrt{f_1}\left( \frac{{\cal Q}' }{ {\cal Q} } 
  (r^2\partial_\tau-\partial_\sigma) - 4r\partial_\tau\right)
  + U_1 \nonumber\\
&\qquad \qquad 
  + \frac{1}{{\cal P}}(p^2 \partial_\tau + \partial_\sigma)^2
  - is \sqrt{f_2}\left( \frac{{\cal P}' }{ {\cal P} } 
  (p^2\partial_\tau+\partial_\sigma) - 4p\partial_\tau\right) 
  + U_2 \Bigg]\tilde{\Phi} = 0 \,, \label{Debyeeq_s}
\end{align}
where
\begin{align}
 U_1 =& \frac{(s+1)(2s+1)}{6}f_1{\cal Q}'' 
  + \frac{(s+1)(8s+1)}{12}f_1'{\cal Q}' \nonumber\\
  &- \frac{(s^2-1)\beta_1^2r^8 - (s-1)(4s+1)\beta_0\beta_1r^6
  +(10s^2-18s-1)\beta_1\beta_2r^4+9s\beta_0\beta_2r^2
  -3s(s+2)\beta_2^2}{3r^6f_1}{\cal Q} \,, \\
 U_2 =& \frac{2s^2+1}{6}f_2{\cal P}''
 + \frac{(2s^2+1)}{12}f_2'{\cal P}'
 - \frac{s^2f_2{\cal P}'^2}{4{\cal P}}
 + \frac{(4s^2+1)\beta_1}{3}{\cal P} \,.
\end{align}
Thus the Debye equations for $\tilde{{\bm h}}^{(1)}$
and $\tilde{{\bm h}}^{(2)}$ are separable.
Setting the separated form
\begin{align}
 \tilde{\Phi} = e^{i\omega\tau}e^{im\sigma}R(r)S(p) \,,
\end{align}
we obtain the ordinary differential equations
\begin{align}
 \frac{1}{\sqrt{f_1}^{2s-1}{\cal Q}^s}\frac{d}{dr}
 \left(\sqrt{f_1}^{2s+1}{\cal Q}^{s+1}\frac{dR}{dr}\right) 
 + V_1 R &= 0 \,, \label{sep_Debye_1} \\
 \sqrt{f_2}\frac{d}{dp}\left(\sqrt{f_2}{\cal P}\frac{dS}{dp}\right) 
 + V_2 S &= 0 \,, \label{sep_Debye_2}
\end{align}
where
\begin{align}
 V_1 =&\frac{1}{{\cal Q}}(\omega r^2 - m)^2
  + is \sqrt{f_1}\left( \frac{{\cal Q}' }{ {\cal Q} } 
  (\omega r^2-m) - 4\omega r\right)
  + \frac{(s+1)(2s+1)}{6}f_1{\cal Q}'' 
  + \frac{(s+1)(8s+1)}{12}f_1'{\cal Q}' \nonumber\\
  &- \frac{(s^2-1)\beta_1^2r^8 - (s-1)(4s+1)\beta_0\beta_1r^6
  +(10s^2-18s-1)\beta_1\beta_2r^4+9s\beta_0\beta_2r^2
  -3s(s+2)\beta_2^2}{3r^6f_1}{\cal Q}
  -\kappa \,, \\
V_2 =& -\frac{(s\sqrt{f_2}{\cal P}'-2\omega p^2-2m)^2}{4{\cal P}}
 - 4s \omega p \sqrt{f_2} 
 + \frac{2s^2+1}{6}f_2{\cal P}''
 + \frac{(2s^2+1)}{12}f_2'{\cal P}'
 + \frac{(4s^2+1)\beta_1}{3}{\cal P}+ \kappa \,.
\end{align}
It is important that
the Debye equations obtained here reduce to the Teukolsky equation
\cite{Teukolsky:1972,Teukolsky:1973} in the limit to the Ricci-flat case, 
where the Wahlquist spacetime becomes the Kerr spacetime.

Since $\tilde{{\bm h}}^{(3)}$ is an ordinary CKY tensor,
the corresponding gauge field $\tilde{{\bm A}}^{(3)}$ is vanishing.
Thus, the Debye equation for $\tilde{{\bm h}}^{(3)}$ is given by
\begin{align}
 \left[g^{ab}\nabla_a\nabla_b
 -\frac{R}{6}+2\Psi_2 \right]\tilde{\Phi} = 0 \,,
\end{align}
which is not separable.
One might think that there could be a gauge transformation that
transforms the Debye equation for ${\bm h}^{(3)}$ to a separable one.
However, we can show that such a gauge transformation does not exist.
Thus, the Debye equation for ${\bm h}^{(3)}$ cannot be solved 
by separation of variables, up to any gauge transformation.

\subsection{Separation constant and symmetry operator}
\label{sec:3-3}

Equation \eqref{Debyeeq_s} is written as
\begin{align}
 {\cal H}\Phi \equiv 
 \frac{1}{r^2+p^2} \left( {\cal S}_1+{\cal S}_2 \right) \Phi = 0 \,,
\end{align}
where
\begin{align}
 {\cal S}_1 =&\frac{1}{\sqrt{f_1}^{2s-1}{\cal Q}^s}\partial_r
 \left(\sqrt{f_1}^{2s+1}{\cal Q}^{s+1}\partial_r\right)
  - \frac{1}{{\cal Q}}(r^2 \partial_\tau - \partial_\sigma)^2 
 +s \sqrt{f_1}\left( \frac{{\cal Q}' }{ {\cal Q} } 
  (r^2\partial_\tau-\partial_\sigma) - 4r\partial_\tau\right)+ U_1 \,, \\
 {\cal S}_2 =& \sqrt{f_2}\partial_p
 \left(\sqrt{f_2}{\cal P}\partial_p\right) 
 + \frac{1}{{\cal P}}(p^2 \partial_\tau + \partial_\sigma)^2
  - is \sqrt{f_2}\left( \frac{{\cal P}' }{ {\cal P} } 
  (p^2\partial_\tau+\partial_\sigma) - 4p\partial_\tau\right)
 + U_2\,.
\end{align}

The separability of \eqref{Debyeeq_s} implies
the existence of a symmetry operator ${\cal K}$
which commutes with ${\cal H}$, i.e., $[{\cal H},{\cal K}]=0$,
because the separation constant is provided as the eigenvalue
of the symmetry operator,
\begin{align}
 {\cal K}\Phi = \kappa \Phi \,.
 \label{EE_K}
\end{align}
The symmetry operator ${\cal K}$ is actually obtained as
\begin{align}
 {\cal K} =
 \frac{1}{r^2+p^2}\left(p^2{\cal S}_1-r^2{\cal S}_2\right) \,.
\end{align}
A direct calculation shows that
the symmetry operator ${\cal K}$ commutes with ${\cal H}$.
From Eqs.~\eqref{sep_Debye_1} and \eqref{sep_Debye_2},
we have ${\cal S}_1\Phi = \kappa \Phi$ and ${\cal S}_2\Phi = -\kappa \Phi$,
and hence we can confirm the relation \eqref{EE_K}.

\subsection{Gauge potentials of Maxwell field}
\label{sec:3-4}

By using \eqref{gp_reconstruction}, we can reconstruct the gauge potentials
of Maxwell field. Now that we have two sets of GCKY tensors
${\bm h}^{(i)}$ and $\tilde{{\bm h}}^{(i)}$,
the corresponding gauge potentials are given by
\begin{align}
 {\cal A}^{(i)}_a
 &= - \nabla^b\left(\Phi^{(i)} h^{(i)}_{ba}\right)
 + \frac{2}{3}\Phi^{(i)}\left(\nabla^b+A^{(i)b}\right)h^{(i)}_{ba} \,, 
 \nonumber\\
 &= - \nabla^b\left(\tilde{\Phi}^{(i)} \tilde{h}^{(i)}_{ba}\right)
 + \frac{2}{3}\tilde{\Phi}^{(i)}\left(\nabla^b
 + \tilde{A}^{(i)b}\right)\tilde{h}^{(i)}_{ba} \,,
\end{align}
where $\Phi^{(i)}$ and $\tilde{\Phi}^{(i)}$ are solutions
to the Debye equations for ${\bm h}^{(i)}$ and $\tilde{{\bm h}}^{(i)}$, respectively.

Finally, we comment its relation to the work by Araneda 
\cite{Araneda:2016, Araneda:2018}.
Since we find that for all $i=1,2,3$,
${\bm h}^{(i)}$ satisfies the relation
\begin{align}
 \left(\nabla^b + A^{(i)b}\right)h^{(i)}_{ba}
 = 3 A^{(3)b}h^{(i)}_{ba} \,,
\end{align}
using which the gauge potentials $\bm{\mathcal A}^{(i)}$ 
can be expressed in an alternative form
\begin{align}
 {\cal A}^{(i)}_a
 = - \left(\nabla^b-2A^{(3)b}\right)
 \left(\Phi^{(i)} h^{(i)}_{ba}\right)\,.
 \label{Aranedaform}
\end{align}
This form of the gauge potentials was pointed out in
\cite{Araneda:2016,Araneda:2018}, where Araneda 
derived the form of gauge potential
on a type-D vacuum spacetime with cosmological constant.
Our result generalizes it to non-vacuum case.
From the viewpoint of gauge invariance,
Araneda's form \eqref{Aranedaform} looks like breaking
the gauge invariance \eqref{ginvariance_A}
since ${\bm A}^{(3)}$ enters in the gauge-covariant derivative.

\section{Summary and discussion}
\label{sec:4}

We have examined hidden symmetry and 
its relation to the separability of the Maxwell equation
on the Wahlquist spacetime.
When the metric is off-shell, that is, $f_1$ and $f_2$ are arbitrary,
the Wahlquist spacetime is of type I rather than type D.
Nevertheless, the PNDs ${\bm k}$ and ${\bm l}$
on this spacetime are shear-free and geodesic,
and hence those shear-free PNDs provide us three GCKY tensors
${\bm h}^{(1)}$, ${\bm h}^{(2)}$, and ${\bm h}^{(3)}$.
Moreover, since ${\bm h}^{(3)}$ satisfies the eigenequation
\eqref{EigenEquation}, the Maxwell equation 
reduces to the Debye equation \eqref{Debye_3}.
When $f_1$ and $f_2$ are given by \eqref{Def_f1f2},
${\bm k}$ and ${\bm l}$ become repeated PNDs,
and hence the Wahlquist spacetime is of type D.
Then, ${\bm h}^{(1)}$ and ${\bm h}^{(2)}$ satisfy
the eigenequation \eqref{EigenEquation},
and we obtain two additional Debye equations,
although the Debye equations for ${\bm h}^{(1)}$ and ${\bm h}^{(2)}$
are not separable.

In order to obtain the separable Debye equation \eqref{Debyeeq_s0},
we have performed the gauge transformations 
\eqref{gtransf_h3}, \eqref{gtransf_h1} and \eqref{gtransf_h2},
keeping the relation ${\bm A}^{(1)}+{\bm A}^{(2)}=2{\bm A}^{(3)}$.
In fact, the Debye equations for ${\bm h}^{(1)}$ and ${\bm h}^{(2)}$
are not separable; on the other hand, the Debye equations for
$\tilde{{\bm h}}^{(1)}$ and $\tilde{{\bm h}}^{(2)}$ are separable.
At this moment, we do not have a clear understanding why the separation of variables is achieved by these gauge transformations.
An observation we can make is that thanks to ${\bm A}^{(3)}$ being pure gauge,
we fixed the gauge transformation \eqref{gtransf_h2}
so that $\tilde{{\bm A}}^{(2)} = - \tilde{{\bm A}}^{(1)}$,
and then the Debye equation for $\tilde{{\bm h}}^{(2)}$ became separable.
Another observation is that in the Kerr-NUT-(A)dS spacetime,
since we have $\tilde{{\bm h}}^{(1)}={\bm h}^{(1)}$ and 
$\tilde{{\bm h}}^{(2)}=\chi^2{\bm h}^{(2)}$,
the Debye equation for ${\bm h}^{(1)}$ is separable
without any gauge transformation.
It would be interesting to investigate 
the origin of the separable property based on these observations.

Recall that the Debye equation \eqref{Debyeeq_s0}
reduces to the Teukolsky equation with $s=0$, $\pm 1/2$, $\pm 1$ and $\pm 2$
in the limit to the Ricci-flat spacetime, i.e., the Kerr spacetime.
This motivates us to ask if the Debye equation \eqref{Debyeeq_s0}
derived on non-vacuum spacetime
makes sense for values of $s$ other than $s=\pm 1$.
In appendix~\ref{app:Dirac} we examine the $s=\pm 1/2$ case, 
and find that the master equations of the massless Dirac equation 
can be derived based on the method of \cite{Benn:1997}.
In \cite{Hinoue:2014}, it was shown that
the torsional Dirac equation on the Wahlquist spacetime
can be solved by separation of variables.
Namely, on the Wahlquist spacetime,
both ordinary and torsional Dirac equations are separable
at least in the massless case.

For $s=\pm 2$, even in four dimensions,
the separation of variables in the linearised Einstein equation
has not been realized so far except in the vacuum case.
It is remarkable that Araneda \cite{Araneda:2016,Araneda:2018}
provided the master equations for the linearized Einstein equation
on a type-D vacuum spacetime with cosmological constant.
The Debye equation (60) for s=2 coincides with his master
equation in the vaccum limit.

For $s=0$, the Debye equation (60) becomes
the conformally invariant equation,
called the conformal-Laplace equation,
\begin{align}
 \left( \Box - \frac{R}{6} \right) \Phi = 0 \,,
\end{align}
which implies that the conformal-Laplace equation
on the conformal class of the type-D Wahlquist spacetime
can be solved by separation of variables.
This seems natural because GCKY symmetry preserves
under any conformal transformation ${\bm g}\to \tilde{{\bm g}}=\Omega^2{\bm g}$.

We have connected the GCKY tensors ${\bm h}^{(\pm)}$
with gauge fields ${\bm A}^{(\pm)}$
to the TCKY tensors ${\bm h}^{(\pm)}$ with torsions 
${\bm T}^{(\pm)}=2i*{\bm A}^{(\pm)}$.
Such a connection between gauged and torsional CKY tensors
is of great interest in the study of hidden symmetry
since there is a little difference between their properties.
For example, we will be able to investigate (gauged) CKY symmetry 
by means of torsional CKY symmetry,
as we have constructed the ordinary CKY tensor $\tilde{{\bm h}}^{(3)}$
on the Wahlquist spacetime in Sec.~\ref{sec:2}.
On the other hand, 
such a connection 
might exist only in four dimensions,
since the torsion and the gauge field are connected by the Hodge dual.

The method used in this paper seems to work only in four dimensions, 
so we need alternative methods to examine the higher-dimensional case~\cite{Hinoue:2014}.
One possibility is to apply the novel method recently developed
for the Kerr-NUT-(A)dS metric in general dimensions 
\cite{Lunin:2017,Frolov:2018,Krtous:2018}, 
which is based on the ordinary CCKY tensor.
The Wahlquist spacetime in general dimensions
admits TCCKY \cite{Hinoue:2014}, 
which suggests that this method works in a parallel manner even in general dimensions. 
However, the existence of torsion could be obstacle for this method.
In \cite{Kubiznak:2019}, this method was applied on the Cveti\v{c}-L\"u-Page-Pope spacetime
in five dimensions, where the Maxwell equation is modified by adding the Chern-Simons term that arises naturally in
the equation of motion in the supergravity theory,
and the additional 
term is realized by means of the torsion.
For now, we have  no reason on the Wahlquist spacetime
to modify the Maxwell equation by torsion,
but it would be fruitful to pursue this issue and clarify the applicability of this method.

\section*{Acknowledgments}
This work was partly supported by Osaka City University Advanced Mathematical Institute
 (MEXT Joint Usage/Research Center on Mathematics and Theoretical Physics).
N.T. is supported by Grant-in-Aid for Scientific Research from the Ministry of Education, 
Culture, Sports, Science and Technology, Japan No.18K03623
and AY 2018 Qdai-jump Research Program of Kyushu University.
Y.Y. is supported by Grant-in-Aid for Scientific Research from the Ministry of Education, 
Culture, Sports, Science and Technology, Japan No.16K05332.

\appendix

\section{Separability of the massless Dirac equation}
\label{app:Dirac}

In this appendix, we sketch the outline of how to obtain the master equations
of the massless Dirac equation
\begin{align}
 \gamma^a \nabla_a \Psi = 0 \,,
 \label{Dirac_eq}
\end{align}
where $\{\gamma^a\}=\{\gamma^0,\gamma^1,\gamma^2,\gamma^3\}$
are the gamma matrices.
In what follows, we use the representations
of the gamma matrices
\begin{align}
&\gamma^0=\left(\begin{array}{cc}
0 & -I \\
I & 0  
\end{array}\right) \,, \quad
\gamma^1=\left(\begin{array}{cc}
0 & I \\
I & 0  
\end{array}\right) \,, \quad
\gamma^2=\left(\begin{array}{cc}
\sigma^1 & 0 \\
0 & -\sigma^1  
\end{array}\right) \,, \quad
\gamma^3=\left(\begin{array}{cc}
\sigma^2 & 0 \\
0 & -\sigma^2 
\end{array}\right) \,,
\end{align}
and, as usual we define $\gamma^5$ as
\begin{align}
 \gamma^5=\gamma^0 \gamma^1 \gamma^2 \gamma^3
 =\left(\begin{array}{cc}
-i \sigma^3 & 0 \\
0 & i \sigma^3
\end{array}\right) \,,
\end{align}
where $\sigma^1$, $\sigma^2$ and $\sigma^3$
are Pauli's matrices.

\subsection{Gauged twistor spinor and the massless Dirac equation}
Benn, Charlton and Kress \cite{Benn:1997} showed that 
if a spacetime admits a gauged twistor spinor\footnote{
See also \cite{Ertem:2017,Ertem:2018} for the integrability conditions
and symmetry operators of the gauged twistor spinor equation
in arbitrary dimensions.}
$\psi$, defined by 
\begin{equation}\label{twist}
 \nabla^{(A)}_a \psi 
 = \frac{1}{4} \gamma_a \gamma^b \nabla_b^{(A)} \psi \,,
\end{equation}
where $\nabla^{(A)}_a$ is the gauge-covariant derivative given by
$\nabla^{(A)}_a = \nabla_a+A_a$,
and if it satisfies the equations
\begin{align}
 \frac{1}{2}F_{ab}\gamma^a \gamma^b \psi = \lambda_F \psi \,,
\label{eigen}
\end{align}
with some function $\lambda_F$ and
the field strength ${\bm F}$ of the gauge field ${\bm A}$,
then the massless Dirac equation reduces to the scalar-type equation
\begin{equation}
\left[g^{ab}\left(\nabla_a-A_a\right)
\left(\nabla_b-A_b\right)+\lambda \right]\Phi=0
\label{Debye-Dirac}
\end{equation}
with
\begin{equation}
 \lambda = -\frac{R}{6} - \frac{\lambda_F}{3} \,.
\end{equation}
Given a solution $\Phi$ to \eqref{Debye-Dirac},
we can reconstruct a solution $\Psi$ to the Dirac equation \eqref{Dirac_eq} by
\begin{equation}
 \Psi = \gamma^a \nabla_a(\Phi\psi)
 -\frac{1}{2}\Phi \gamma^a \nabla^{(A)}_a \psi \,.
 \label{Dirac-reconst}
\end{equation}
The chilarity of this Dirac field $\Psi$ becomes opposite
to the chirality of the gauged twistor spinor $\psi$.

\subsection{Dirac equation on the Wahlquist spacetime}
On the off-shell Wahlquist spacetime, we obtain the gauged twistor spinors as
\begin{equation}
\psi_-^{(1)}=e^{-i \alpha}\left(\begin{array}{c}
0 \\
u \\
0\\
0
\end{array}\right) \,, \quad
\psi_+^{(1)}=e^{i \alpha}\left(\begin{array}{c}
u \\
0\\
0\\
0
\end{array}\right) \,, \quad
\psi_-^{(2)}=e^{i \alpha}\left(\begin{array}{c}
0 \\
0 \\
v\\
0
\end{array}\right)\,, \quad
\psi_+^{(2)}=e^{-i \alpha}\left(\begin{array}{c}
0 \\
0\\
0\\
v
\end{array}\right) \,, \label{GTSpinor}
\end{equation}
where
\begin{equation}
 u =\left(\frac{r^2+p^2}{Q f_1}\right)^{1/4} \,, \quad
 v = \frac{1}{\sqrt{2}}\left(\frac{Q f_1}{r^2+p^2}\right)^{1/4} \,, \quad
 \alpha = \frac{1}{2} \tan^{-1} \left( \frac{r}{p} \right) \,.
\end{equation}
The signs $\pm$ 
correspond to the chirality as
\begin{equation}
i \gamma^5 \psi_\pm^{(1)}=\pm \psi_\pm^{(1)} \,, \quad
i \gamma^5 \psi_\pm^{(2)}=\pm \psi_\pm^{(2)}\,.
\end{equation}
The gauge fields for $\psi_-^{(1)}$
and $\psi_-^{(2)}$ are given by $\frac{1}{2}{\bm A}^{(1)}$ and 
$\frac{1}{2}{\bm A}^{(2)}$,
where ${\bm A}^{(1)}$ and ${\bm A}^{(2)}$ are the gauge fields for 
GCKY tensors given by \eqref{gp_gcky_1} and \eqref{gp_gcky_2},
and the gauge fields for $\psi_+^{(1)}$ and $\psi_+^{(2)}$
are given by the complex conjugates of the gauge fields 
for $\psi_-^{(1)}$ and $\psi_-^{(2)}$, respectively.
These twistor spinors are related to the PNDs by 
\begin{align}
 {\bm \ell}_* 
 = \left(\bar{\psi}_-^{(1)} \gamma_a \psi_-^{(1)}\right) \bm{e}^a
 = \left(\bar{\psi}_+^{(1)} \gamma_a \psi_+^{(1)}\right) \bm{e}^a \,, \quad
 \bm{k}_*
 = \left(\bar{\psi}_-^{(2)} \gamma_a \psi_-^{(2)}\right) \bm{e}^a 
 = \left(\bar{\psi}_+^{(2)} \gamma_a \psi_+^{(2)}\right) \bm{e}^a \,.
\end{align}
It is shown that these twistor spinors satisfy Eq.~\eqref{eigen}
with 
\begin{align}
& \lambda = -\frac{R}{6} - \Psi_2 \qquad
  \textrm{for $\psi_-^{(1)}$, $\psi_-^{(2)}$} \,, \\
& \lambda = -\frac{R}{6} - \bar{\Psi}_2 \qquad
  \textrm{for $\psi_+^{(1)}$, $\psi_+^{(2)}$} \,,
\end{align}
where $\bar{\Psi}_2$ is the complex conjugate of $\Psi_2$.
Thus, after the gauge transformations \eqref{gtransf_h1}
and \eqref{gtransf_h2} for the gauge fields, we obtain
\begin{align}
 \left[g^{ab}\left(\nabla_a \pm \frac{1}{2}\tilde{A}^{(1)}_a\right)
 \left(\nabla_b \pm \frac{1}{2}\tilde{A}^{(1)}_b\right)
  - \frac{R}{6} - \Psi_2\right]\tilde{\Phi} = 0 \,,
\end{align}
where the plus and minus signs in front of $\frac{1}{2}$
 correspond to the cases using $\psi_-^{(2)}$ and $\psi_-^{(1)}$, respectively.
An important consequence is that this equation coincides with 
Eq.~\eqref{Debyeeq_s0} with $s=\pm 1/2$, and hence this can be solved by separation of variables.

Finally, we remark that the gauged twistor spinors \eqref{GTSpinor}
give rise to three GCKY tensors
\begin{align}
 {\bm h}^{(1)}
 =& \left(\bar{\psi}^{(1)}_+[\gamma_a,\gamma_b]\psi^{(1)}_-\right)
 \,\bm{e}^a\wedge\bm{e}^b \,, \\
 {\bm h}^{(2)}
 =& \left(\bar{\psi}^{(2)}_+[\gamma_a,\gamma_b]\psi^{(2)}_-\right)
 \,\bm{e}^a\wedge\bm{e}^b \,, \\
 {\bm h}^{(3)}
 =& \left(\bar{\psi}^{(1)}_+[\gamma_a,\gamma_b]\psi^{(2)}_-\right)
 \,\bm{e}^a\wedge\bm{e}^b \,,
\end{align}
with the gauge fields ${\bm A}^{(1)}$, ${\bm A}^{(2)}$
and ${\bm A}^{(3)}=\frac{1}{2}\left({\bm A}^{(1)}+{\bm A}^{(2)}\right)$.
These GCKY tensors coincide with the GCKY tensors
given by \eqref{Def_h1}--\eqref{Def_h3}
up to some normalisation constants.


\begin{thebibliography}{99}
 
 \bibitem{Teukolsky:1972}
 S. A. Teukolsky,
 Rotating black holes: Separable wave equations for gravitational 
 and electromagnetic perturbations,
 Phys. Rev. Lett. 29 (1972) 1114.
 
 \bibitem{Teukolsky:1973}
 S. A. Teukolsky,
 Perturbations of a rotating black hole.
 1. Fundamental equations for gravitational,
 electromagnetic and neutrino field perturbations,
 Astrophys. J. 185 (1973) 635.
 
 \bibitem{Cohen:1974}
 J. M. Cohen, L. S. Kegeles,
 Electromagnetic fields in curved spaces: A constructive procedure, 
 Phys. Rev. D 10 (1974) 1070-1084.
 
 \bibitem{Cohen:1979}
 J. M. Cohen, L. S. Kegeles,
 Constructive procedure for perturbations of spacetimes,
 Phys. Rev. D 19 (1979) 1641-1664.
 
 \bibitem{Benn:1997}
 I. M. Benn, P. Charlton, J. Kress,
 Debye Potentials for Maxwell and Dirac Fields 
 from a Generalisation of the Killing-Yano Equation,
 J. Math. Phys. 38 (1997) 4504-4527.
 [arXiv:gr-qc/9610037]
 
 \bibitem{Wahlquist:1968}
 H. D. Wahlquist,
 Phys. Rev., {\bf 172} (1968) 1291.
 
 \bibitem{Kramer:1985}
 D. Kramer,
 Perfect fluids with vanishing Simon tensor,
 Class. Quantum Grav., {\bf 2} (1985) L135-L139.
 
 \bibitem{Senovilla:1987}
 J. M. M. Senovilla,
 Stationary axisymmetric perfect-fluid metric with $q+3p=$const,
 Phys. Lett., {\bf A123} (1987) 211-214.
 
 \bibitem{Wahlquist:1992}
 H. D. Wahlquist,
 The problem of exact solutions for rotating rigid bodies in general relativity
 J. Math. Phys., {\bf 33} (1992) 304-335; Erratum {\bf 33} (1992) 3255.
 
 \bibitem{Mars:2001}
 M. Mars,
 The Wahlquist-Newman solution,
 Phys. Rev. {\bf D63} (2001) 064022.
 [arXiv:gr-qc/0101021]
 
 \bibitem{Bradley:1999}
 M. Bradley, G. Fodor, M. Marklund, Z. Perj\'es,
 The Wahlquist metric cannot describe an isolated rotating body
 Class. Quantum Grav. {\bf 17} (2000) 351-359.
 [arXiv:gr-qc/9910001]
 
 \bibitem{Hinoue:2014}
 K. Hinoue, T. Houri, C. Rugina, Y. Yasui,
 General Wahlquist metric in all dimensions,
 Phys. Rev. D 90 (2014) 024037
 [arXiv:1402.6904[gr-qc]]
 
 \bibitem{Houri:2012}
 T. Houri, D. Kubiznak, C. M. Warnick, Y. Yasui,
 Local metrics admitting a principal Killing-Yano tensor with torsion,
 Class. Quantum Grav. 29 (2012) 165001
 [arXiv:1203.0393[hep-th]]
 
 \bibitem{Houri:2019}
 T. Houri, N. Tanahashi, Y. Yasui,
 On symmetry operators for the Maxwell equation
 on the Kerr-NUT-(A)dS spacetime
 [arXiv:1908.10250[gr-qc]]
 
 \bibitem{Araneda:2016}
 B. Araneda,
 Symmetry operators and decoupled equations for linear fields
 on black hole spacetimes,
 Class. Quantum Grav. 34 (2017) 035002
 [arXiv:1610.00736[gr-qc]]
 
 \bibitem{Araneda:2018}
 B. Araneda,
 Generalized wave operators, weighted Killing fields, 
 and perturbations of higher dimensional spacetimes,
 Class. Quantum Grav. 35 (2018) 075015
 [arXiv:1711.09872 [gr-qc]]
 
 \bibitem{Lunin:2017}
 O. Lunin,
 Maxwell's equations in the Myers-Perry geometry,
 JHEP 12 (2017) 138,
 [arXiv:1708.06766]
 
 \bibitem{Frolov:2018}
 V. P. Frolov, P. Krtou\v{s}, D. Kubiz\v{n}\'ak,
 Separation of variables in Maxwell equations
 in Plebanski-Demianski spacetime,
 Phys. Rev. D 97 (2018) 101701,
 [arXiv:1802.09491[hep-th]]
 
 \bibitem{Krtous:2018}
 P. Krtou\v{s}, V. P. Frolov, D. Kubiz\v{n}\'ak,
 Separation of Maxwell equations in Kerr-NUT-(A)dS spacetimes,
 Nucl. Phys. B 934 (2018) 7-38,
 [arXiv:1803.02485[hep-th]]
 
 \bibitem{Kubiznak:2019}
 R. Cayuso, F. Gray, D. Kubiznak, A. Margalit, 
 R. G. Souza, L. Thiele,
 Principal Tensor Strikes Again: 
 Separability of Vector Equations with Torsion,
 [arXiv:1906.10072[hep-th]]
 
 \bibitem{Ertem:2017}
 \"U. Ertem,
 Gauged twistor spinors and symmetry operators,
 J. Math. Phys. 58 (2017) 032302
 [arXiv:1610.02510[hep-th]]
 
 \bibitem{Ertem:2018}
 \"U. Ertem,
 Spin Geometry and Some Applications,
 [arXiv:1801.06988[math-ph]]
 
 
\end{thebibliography}
\end{document}